\documentclass[twocolumn,showpacs,preprintnumbers]{revtex4}
\usepackage{amssymb}
\usepackage{amsfonts}
\usepackage{amsmath}
\usepackage{graphicx}
\usepackage{dcolumn}
\usepackage{bm}

\input{tcilatex}

\begin{document}

\title{Scaling flow diagram in the fractional quantum Hall regime of
GaAs/AlGaAs heterostructures}
\author{S. S. Murzin$^{1}$, S. I. Dorozhkin $^{1}$, D. K. Maude$^{2}$, and
A. G. M. Jansen$^{\text{3}}$}
\affiliation{$^1$ Institute of Solid State Physics RAS, 142432, Chernogolovka, Moscow
District, Russia \\
$^{2}$ Grenoble High Magnetic Field Laboratory, Max-Planck-Institut f\"{u}r
Festk\"{o}rperforschung and Centre National de la Recherche Scientifique, BP
166, F-38042, Grenoble Cedex 9, France\\
$^{\text{3}}$Service de Physique Statistique, Magn\'{e}tisme, et
Supraconductivit\'{e}, D\'{e}partement de Recherche \\
Fondamentale sur la Mati\`{e}re Condens\'{e}e, CEA-Grenoble, 38054 Grenoble
Cedex 9, France }

\begin{abstract}
The temperature driven flow lines of the Hall and dissipative
magnetoconductance data $(\sigma _{xy},\sigma _{xx})$ are studied in the
fractional quantum Hall regime for a 2D electron system in GaAs/Al$_{x}$Ga$%
_{1-x}$As heterostructures. The flow lines are rather well described by a
recent unified scaling theory developed for both the integer and the
fractional quantum Hall effect in a totally spin-polarized 2D electron
system which predicts that one ($\sigma _{xy},\sigma _{xx}$) point
determines a complete flow line.
\end{abstract}

\maketitle

The scaling treatment was initially proposed for the integer quantum Hall
effect (QHE) using a graphical representation of the magnetoconductance data
in the form of the flow diagram \cite{Pr,Khm,Prb}. The flow diagram depicts
the coupled evolution of the diagonal ($\sigma _{xx}$) and Hall ($\sigma
_{xy}$) conductivity components due to diffusive interference effects. The
similarity of many features at different integer and fractional quantum Hall
states stimulated the creation of unified scaling theories \cite%
{Lut,Dolan,BDD,BD,Pr99}. The results of these theories are in qualitative
agreement with each other, however, the phenomenological approach \cite%
{Lut,Dolan,BDD,BD} provides a quantitative picture, the experimental
verification of which is the goal of our investigation.

In Ref.\cite{Lut}, the conjecture was made that the scaling flow diagram for
a totally spin-polarized electron system is invariant under the modular
transformation 
\begin{equation}
\tilde{\sigma}\leftrightarrow \frac{a\sigma +b}{2c\sigma +d},  \label{mod}
\end{equation}%
for complex conductivity $\sigma =\sigma _{xy}+i\sigma _{xx}$ where $a,$ $b,$
$c$, and $d$ are integers with $ad-2bc=1$. This means that any part of the
flow diagram in the ($\sigma _{xy}$,$\sigma _{xx}$) plane can be obtained
from any other part by transformation (\ref{mod}) with corresponding choice
of $a,$ $b,$ $c$, and $d$. Such transformations can be considered \cite{BDD}
as an extension of the \textquotedblleft law of corresponding
states\textquotedblright\ describing the symmetry relations between the
different QHE phases \cite{KLZ}. For a holomorphic scaling $\beta -$function
in the variable $\sigma _{xy}+i\sigma _{xx}$ exact expressions for the flow
lines have been derived \cite{Dolan}.

\begin{figure}[tb]
\includegraphics[width=8cm,clip]{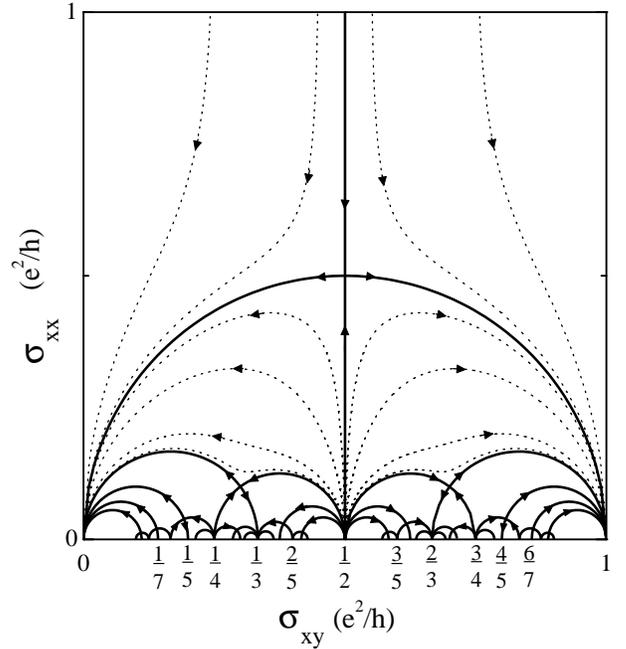}
\caption{The scaling flow diagram showing the coupled evolution of the
diagonal ($\protect\sigma _{xx}$) and Hall ($\protect\sigma_{xy}$)
conductivity components with increasing coherence length for a totally
spin-polarized 2D electron system in the range $0<\protect\sigma _{xy}<1$.
Solid lines represent the semicircle separatrixes and dotted lines give
examples of some other flow lines. }
\label{flow}
\end{figure}

The structure of the theoretical flow diagram \cite{Dolan}, periodic along
the $\sigma _{xy}$-axis, is shown in Fig.\ref{flow} for one period $0<\sigma
_{xy}<1$ ($\sigma _{xx}$ and $\sigma _{xy}$ are in units of$\ e^{2}/h$). For 
$\sigma _{xx}\gg 1$, the points $(\sigma _{xy},\sigma _{xx})$ flow down
vertically with increasing coherence length $L$, deviate from the vertical
at about $\sigma _{xx}\sim 1$, and converge to integer values of $\sigma
_{xy}$ (0 or 1 for the range under consideration) on approaching the
semicircle separatrix 
\begin{equation}
\sigma _{xx}^{2}+\left[ \sigma _{xy}-1/2\right] ^{2}=1/4.  \label{sl1}
\end{equation}%
The points on the vertical separatrix $\sigma _{xy}=1/2$, separating
different QHE phases, flow vertically to the critical point $(1/2,1/2)$ both
from above and from below. Eq.~(\ref{mod}) maps the vertical separatrix at $%
\sigma _{xy}=1/2$ into semicircle separatrixes connecting pairs of points $%
(p_{i}/q_{i},0)$ for odd integer $p_{i},$ and even integer $q_{i}$, with
flow out from these points. The semicircle separatrix Eq.~(\ref{sl1}) is
mapped onto semicircle separatrixes connecting pairs of points $%
(p_{i}/q_{i},0)$ with odd $q_{i}$ and flow direction towards these points
corresponding to either an integer ($q_{i}=1$) or fractional ($q_{i}>1$)
quantum Hall state, or an insulating state ($p_{i}=0$). Crossings of the
different type semicircle separatrixes produce the fixed critical points.
All flow lines different from the separatrixes leave from the points $%
(p_{i}/q_{i},0)$ with even $q_{i}$ and arrive at the points $(p_{i}/q_{i},0)$
with odd $q_{i}$ as it shown in Fig.\ref{flow} by dotted lines below the
large semicircle separatrix of Eq.~(\ref{sl1}). For the sake of clarity,
only a limiting number of semicircle separatixes having the larger diameters
are shown. Since the flow lines do not intersect the separatrixes the latter
divide the flow diagram into the regions where the flow lines can reach only
particular points on the horizonal axis (i.e., particular QHE states).

As follows from the above consideration, the scaling theory \cite%
{Lut,Dolan,BDD,BD} predicts existence of a fractional quantum Hall effect
(FQHE) state at all rational fractions with odd denominators, including the
recently observed \cite{Tsui} the $4/11$- and $5/13$- states. To describe
these states the widely accepted composite fermion picture \cite{Jain} has
to be modified by introducing the concept of higher generation composite
fermions \cite{Jain 04,Frad,Go}.

According to the scaling theory \cite{Dolan}, close to zero temperature,
transitions between QHE states occurring at variation of the magnetic field
or the electron density are described by semicircles connecting
corresponding points on the $(\sigma _{xy},\sigma _{xx})$ diagram. Two such
semicircles have been observed experimentally \cite{Hilke} for the case of
transitions to the insulating state $(0,0)$ from the $(1,0)$ and from the $%
(1/3,0)$ QHE state.

It is worth to note that the flow lines depend on the magnetic field and
such parameters of the system as electron density, disorder, or scattering
mechanism only via the values of the magnetoconductances $\sigma _{xy}$ and $%
\sigma _{xx}$. Where $\sigma _{xy}$ is mainly determined by the electron
density and the magnetic field, $\sigma _{xx}$ is strongly dependent on the
sample quality. For a better sample with higher mobility, the lower $\sigma
_{xx}$ value at given $\sigma _{xy}$ and temperature allows the observation
of a larger number of FQHE states. However, experimental limitations of
attainable low temperatures do not allow to reach the upper parts of the
flow lines in the flow diagram for the high-mobility samples. Therefore, to
investigate the total diagram, samples of very different quality should be
utilized. The data presented here cover that part of the diagram, which
corresponds to samples of moderate quality.

For the integer QHE, the experimentally obtained flow diagram \cite{Wei} was
in qualitative agreement with the theoretical picture. Quantitative
agreement between theoretical \cite{Dolan} and experimental flow lines has
been found for strongly disordered GaAs layers \cite{M02}. To the best of
our knowledge the temperature driven flow diagram in the fractional QHE
regime has been briefly investigated \cite{Cl} for $1<\sigma _{xy}<2$ only.
In this situation, the electron system is not fully spin polarized and the
later appeared scaling theory \cite{Lut} is not directly applicable.
However, in a wide variety of samples \cite{Sh,Wong}, for the transition
from the fractional state $(1/3,0)$ to the insulating state $(0,0)$, the
value of $\rho _{xx}$ at the temperature independent point has been found to
be close to the theoretical critical value $\rho _{xx}^{c}=h/e^{2}$.

\begin{figure}[tb]
\includegraphics[width=8.5cm,clip]{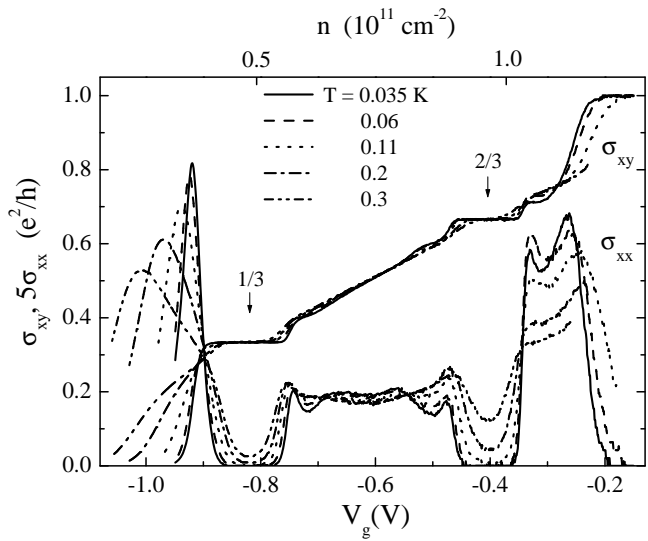}
\caption{The diagonal ($\protect\sigma _{xx}$) and Hall ($\protect\sigma%
_{xy} $) conductivities as a function of the gate voltage for sample 2 at a
6.0 T magnetic field for different temperatures below 0.3~K. The arrows
indicate the gate voltage positions of the $\protect\sigma _{xx}$ minima for
fractional QHE states 2/3 and 1/3.}
\label{G36}
\end{figure}

In the work presented here we explore the temperature driven flow diagram of 
$\sigma _{xx}(T)$ versus $\sigma _{xy}(T)$ for a spin-polarized 2D electron
system in GaAs/A$_{x}$Ga$_{1-x}$As heterostructures in the fractional QHE
regime with $\sigma _{xy}<2/3$. A good agreement with the recent scaling
theories is found, which implies surprisingly that at low enough
temperatures the evolution with temperature of a totally spin-polarized 2D
electron system depends only on initial values of $\sigma _{xy}$ and $\sigma
_{xx}$and that the behavior in different parts of the ($\sigma _{xy},\sigma
_{xx}$) plane is connected by simple transformations according to Eq.~(\ref%
{mod}).

\begin{figure*}[t]
\includegraphics[width=16cm,clip]{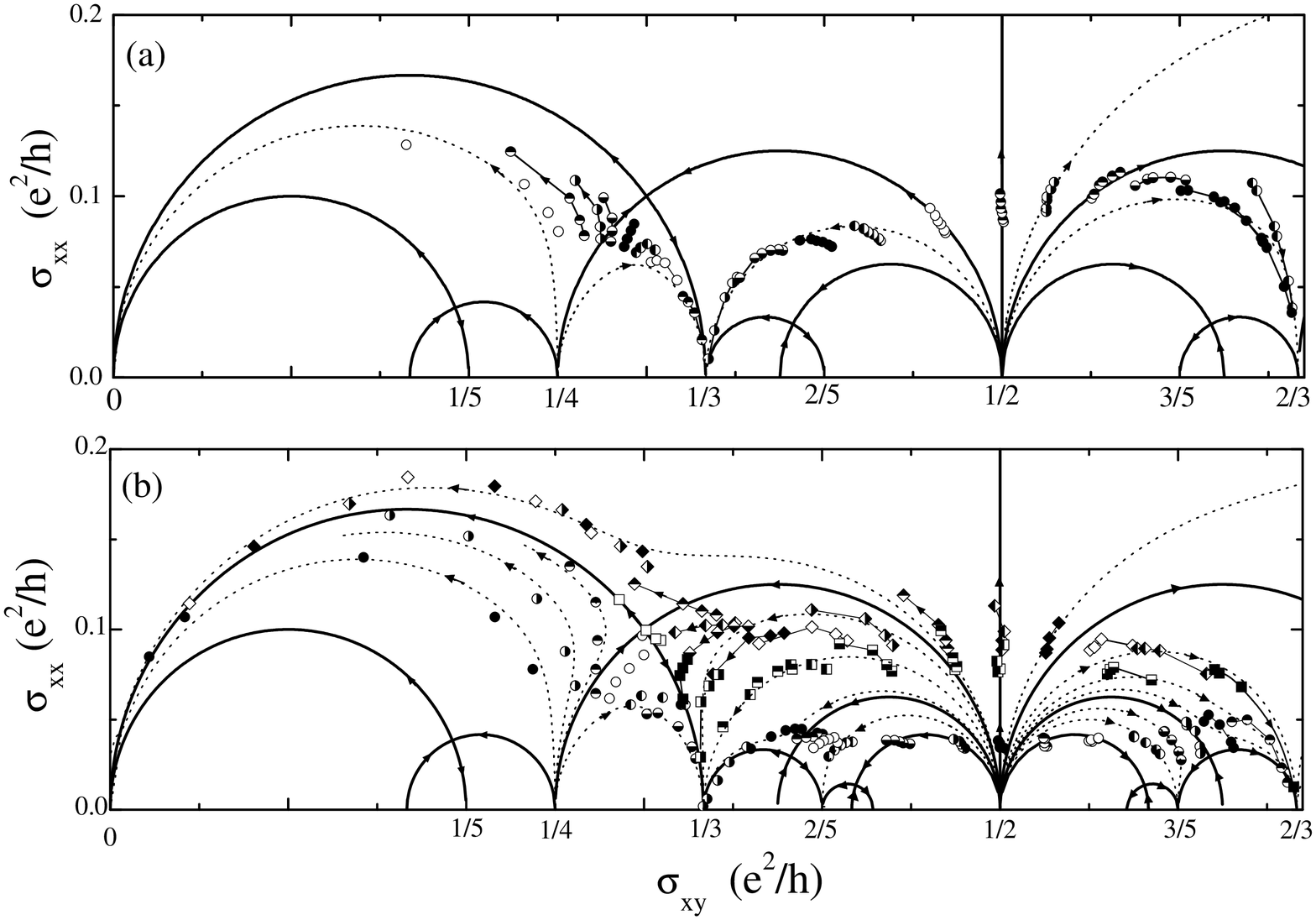}
\caption{Flow-diagram of the temperature dependent ($\protect\sigma _{xy}(T),%
\protect\sigma _{xx}(T)$) data points (a) for sample 1 from 0.37 down to
0.05~K (for $1/3<\protect\sigma _{xy}<2/3)$ and down to $0.11$\ K (for $%
\protect\sigma _{xy}<1/3$) at different magnetic fields in the range $8.5 -
21$~T, and (b) for sample 2 from 0.3 down to 0.035~K at different gate
voltages for $B=2.5$ (diamonds), $2.8$ (squares, a few data), and $6$~T
(circles). Different fillings of the symbols are used for different magnetic
fields (data of sample 1) and for different gate voltages (data of sample
2). Solid lines represent the theoretical vertical and semicircle
separatrixes. Dotted lines show flow lines close to experimental data.}
\label{fl2}
\end{figure*}

Two samples of MBE grown GaAs/Al$_{x}$Ga$_{1-x}$As heterostructures from
different wafers were used in our experiments. They have an undoped Al$_{x}$%
Ga$_{1-x}$As spacer of 50 and 70~nm for sample 1 and 2, respectively. Sample
1, without gate structure, has a low-temperature electron density $%
n=1.35\times 10^{11}$~cm$^{-2}$ and a mobility $8.8\times 10^{4}$~cm$^{2}$
/Vs. Sample 2, with a gate structure, has an electron density $n=1.4\times
10^{11}$~cm$^{-2}$ and a mobility $1.2\times 10^{6}$~cm$^{2}$/Vs at zero
gate voltage. Previous study of sample 2 \cite{Dor2} proved a
spin-depolarization of the 2D electron system at filling factors $2/3<\nu <1$%
, in agreement with theoretical prediction \cite{Ch}. For this reason we
limit the here presented flow diagrams to $0<\sigma _{xy}<2/3$. Note that
because of very different mobilities of these two samples the same values of 
$\sigma _{xy}$ and $\sigma _{xx}$ at the same temperature are obtained at
very different electron densities. For example, the point (0.5,~0.1) for
sample 2 is reached at $n\approx 3\times 10^{10}$\emph{~}cm$^{-2}$, a factor
of four smaller than the carrier density in sample 1. The Hall bar geometry
was used for measurements of the Hall ($\rho _{xy}$) and diagonal ($\rho
_{xx}$) resistivities. For sample 1 without a gate the diagonal ($\rho _{xx}$%
) and Hall ($\rho _{xy}$) resistivities have been measured as a function of
the magnetic field $B$ at different temperatures and converted into the $%
\sigma _{xx}(B)$ and $\sigma _{xy}(B)$ data. For sample 2 with a gate $\rho
_{xx}$ and $\rho _{xy}$ have been measured as a function of the gate voltage 
$V_{g}$ at different temperatures for three fixed magnetic fields $2.5$, $2.8
$and $6$~T. The absolute values of the Hall $\rho _{xy}$ signals were
appreciably different for two opposite directions of the magnetic field at
the highest fields for sample 1 and the lowest gate voltages for sample 2,
probably because of an admixture of $\rho _{xx}$to $\rho _{xy}$. The average
has been taken as $\rho _{xy}$.

The corresponding data for $\sigma _{xx}(V_{g})$ and $\sigma _{xy}(V_{g})$
(both given in units of $e^{2}/h$) are shown in Fig.~\ref{G36} for $B=6$~T. 
\emph{\ }The two fractional QHE states ($2/3$ and $1/3$) are well pronounced
especially at the lowest temperature. These well developed FQHE states are
observed together with weakly pronounced 2/5 and 3/5~states. The upper axis
for the electron density $n$ is slightly nonlinear below $n\approx 3\times
10^{10}$~cm$^{-2}.$ An additional shallow minimum in $\sigma _{xx}$ and a
step in $\sigma _{xy}$ can be seen at $n=1.1\times 10^{11}$~cm$^{-2}$ ($\nu
\approx 0.75$). However, their evolution with temperature differs from the
usual behavior for the fractional QHE. The value of $\sigma _{xx}$ in the
minimum increases with decreasing temperature and the plateau value of $%
\sigma _{xy}$ depends on temperature. This observation is in agreement with
earlier results \cite{Kl} and probably originates from partial spin
polarization of the 2D electron system at filling factors $2/3<\nu <1$ which
makes the theory \cite{Lut,Dolan,BDD,BD,Pr99} inapplicable in this range. At 
$B=2.5$ and $2.8$~T, the fractional QHE is observed only at $\nu =1/3$ and $%
2/3$ as for the data of sample 1.

In Fig.~\ref{fl2}, the experimental flow data are compared with theoretical
flow diagram and rather good agreement is demonstrated. For sample 1 (Fig.~%
\ref{fl2}(a)) at $\sigma _{xy}=1/2$ the flow of the experimental data goes
vertically up in accordance with the theory. To the left of this line the
flow lines deviate to the left and are nearly horizontal around $\sigma
_{xy}=0.4$, before curving down tending to $(1/3,~0)$. Similarly to the
right of $\sigma _{xy}=1/2$ the experimental flow deviates to the right, is
almost horizontal around $\sigma _{xy}=0.6$, and then curves down tending to 
$(2/3,~0)$. At $0.25<\sigma _{xy}<0.3$ the flow lines go approximately up at
the beginning and diverge near the critical point close to the theoretically
predicted one at $(0.3,~0.1)$.

\begin{figure}[t]
\includegraphics[width=7.5cm,clip]{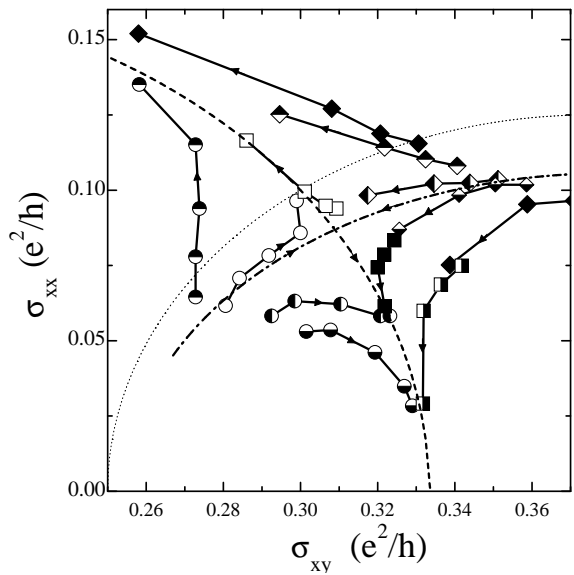}
\caption{A part of the flow diagram of Fig.~\protect\ref{fl2}(b) in the
vicinity of the theoretical critical point $(0.3,0.1)$. Experimental points
(diamonds for $B=2.5$, squares for $2.8$ , and circles for $6$~T ) are
connected by solid lines and the direction of their shift with lowering
temperature are shown by arrows. The theoretical semicircle separatrixes are
presented by dashed and dotted lines. Dash-dotted line is obtained by a
slight deformation of the dotted theoretical line.}
\label{fl4}
\end{figure}

For sample 2 (Fig.~\ref{fl2}(b)), the experimental points reproduce a large
part of the theoretical flow lines connecting points $(1/2,~0)$ and $(1/3,~0)
$; $(1/2,~0)$ and $(2/3,~0)$; $(1/4,~0)$ and $(0,~0)$, and follow vertical
flow line at $\sigma _{xy}=1/2$. The flow line between the $(1/2,~0)$ and $%
(0,~0)$ points also demonstrate very good agreement with the experimental
data. Note that the experimental points shift with lowering temperature
always in the direction of the nearest flow lines in complete agreement with
what is expected from the theory. For example, experimental points move with
lowering temperature along the $\sigma _{xx}$ axis upward in the vicinity of 
$\sigma _{xy}=1/2$ and $\sigma _{xy}=1/4$ and downward in the vicinity of $%
\sigma _{xy}=1/3$ and $\sigma _{xy}=2/3$. The theoretical flow lines
directed to fractional QHE states with $\sigma _{xy}=2/5$ and $\sigma
_{xy}=3/5$ are also rather well confirmed experimentally. A complicated
picture exists in the vicinity of the theoretical critical point $(0.3,~0.1)$%
. Corresponding data are shown in Fig.~\ref{fl4} on an enlarged scale. The
experimental data can be well fitted by slightly deforming the separatrix
connecting the $(1/2,~0)$ and $(1/4,~0)$ points (dotted line) as is shown by
dash-dotted line in Fig.~\ref{fl4}. The flow of the experimental data has a
nontrivial character: the data points shift towards the critical point when
located in the vicinity of the deformed separatrix and away from the
critical point along separatrix connecting $(1/3,~0)$ and $(0,~0)$ points
(dashed line). The separatrix deformation results in shift of the critical
point to the $(0.31,~0.09)$ position. Thus, there is only rather small
quantitative discrepancy (about 10\%) between theoretical and our
experimental flow lines. This difference could be attributed to macroscopic
inhomogeneity of the sample bearing in mind the data scattering for $\rho
_{xx}^{c}$, 30\% in Ref.\cite{Sh} and 10\% in Ref.\cite{Wong}.

The possible role of dissipationless edge currents could have an influence
on the correct measurement of the magnetotransport tensor components in our
experiment. The edge currents can shortcircuit the dissipative bulk
conductivity so that neither magnetoresistivity nor magnetoconductivity can
be determined from the measured magnetoresistance \cite{Tsui87,Dor90}. The
shortcircuiting by edge currents could occur only at filling factors above a
very well pronounced QHE state (or, equivalently, at magnetic fields and
gate voltages, respectively, below and above QHE plateaux) giving rise to an
asymmetric form of the magnetoconductivity $\sigma _{xx}(B)$ around a QHE
minimum. Moreover, in the case of fractional QHE, edge currents exist only
in a rather close vicinity of the corresponding QHE states \cite{Goldman}
(for example, edge currents related to FQHE do not affect the conductivity
at $\nu =1/2$). Considering our data below $\nu =2/3$, pronounced fractional
QHE states occur only at $\nu =1/3$ at the lowest temperatures and highest
fields with $\sigma _{xx}$ approaching zero. Therefore, the data points on
the flow lines converging to the insulating state $(0,0)$, those in the
vicinity of the discussed critical point $(0.31,~0.09)$, and those around
and above $\sigma _{xy}=1/2$ up to 2/3 cannot be affected by the edge
currents. The quite symmetric form of the magnetoconductivity minima at $\nu
=1/3$ observed in our experiment even at the lowest temperatures (see Fig.~%
\ref{G36} as an example) gives a further indication that edge currents
aren't of importance in our data.

The theoretical flow diagram describes the behavior of a 2D system when only
the coherence length depends on temperature with all other parameters being
temperature independent. Such regime is obviously limited from both low and
high temperature sides. At high temperatures, for example, the formation of
an energy gap in the excitation spectrum with decreasing temperature is not
included in the scaling theory. At low temperatures the transition to
variable-range hopping could break the scaling description. However, the
latter would change the flow diagram only in very close vicinity of the
quantum Hall states. An estimation of the upper temperature limit for the
scaling approach is not well established. Existing experimental results on
the temperature dependence of the compressibility of a 2D electron system in
the fractional quantum Hall effect regime~\cite{Dor,Eis} and on the
non-monotonic temperature dependence of $\sigma _{xx}$ at $\sigma _{xy}=1/2$~%
\cite{Rok} imply the upper temperature limit for the scaling flow lines to
be at approximately $0.5$ K. In our samples at temperatures above $0.4$~K
the deviations of the experimental data from the scaling theory predictions
increase gradually, consistently with results of Ref.\cite{Dor,Eis,Rok}.

In summary, for the fractional quantum Hall effect regime at $0<\sigma
_{xy}<2/3$ we have found that the complicated temperature dependence of the
magnetoconductivity tensor components when presented in the form of the flow
lines on the ($\sigma _{xy}$,$~\sigma _{xx}$) plane is rather well
quantitatively described by equations of the scaling theory \cite%
{Lut,Dolan,BDD,BD}. The evolution of $(\sigma _{xy},~\sigma _{xx})$ with
decreasing temperature is determined uniquely by the initial values of $%
\sigma _{xy}$ and $\sigma _{xx}$ independently from the microscopic
parameters of a totally spin-polarized 2D electron system.

The authors greatly appreciate producing of samples by MBE groups from
Institute of Semiconductor Physics RAS, Novosibirsk, Russia (sample 1) and
from Max-Planck-Institut f\"{u}r Festk\"{o}rperforschung, Stuttgart, Germany
(sample 2). This work was supported by the \emph{Russian Foundation for
Basic Research.}


\begin{thebibliography}{99}
\bibitem{Pr} H. Levine, S. B. Libby, and A. M. M. Pruisken, Phys.\ Rev.\
Lett.\ \textbf{51}, 1915 (1983).

\bibitem{Khm} D.E.Khmel$^{\prime }$nitski\u{\i},Pis$^{\prime }$ma Zh.\
Eksp.\ Teor.\ Fiz, \textbf{38}, 454 (1983) [JETP Lett. \textbf{38}, 552
(1983)].

\bibitem{Prb} A. M. M. Pruisken in \emph{The Quantum Hall Effect}, edited by
R. E. Prange and S. M. Girven, Springer-Verlag, 1990.

\bibitem{Lut} C.A. L\"{u}tken and G.G. Ross, Phys. Rev. B \textbf{45}, 11837
(1992); B \textbf{48}, 2500 (1993).

\bibitem{Dolan} B. P. Dolan, Nucl. Phys. B \textbf{554}, 487 (1999); J.
Phys. A \textbf{32}, L243 (1999).

\bibitem{BDD} C.P. Burgess, Rim. Dib, and B.P. Dolan, Phys. Rev. B \textbf{62%
}, 15 359 (2000)..

\bibitem{BD} C. P. Burgess and B. P. Dolan, Phys.\ Rev.\ B \textbf{63},
155309 (2001).

\bibitem{Pr99} A. M. M. Pruisken, M. A. Baranov, and B. \v{S}kori\'{c},
Phys. Rev. B \textbf{60}, 16 807 (1999).

\bibitem{KLZ} S. Kivelson, D. Lee, and S. Zang, Phys.\ Rev.\ B \textbf{46},
2223 (1992).

\bibitem{Hilke} M. Hilke, D. Shahar, S. H. Song, D. C. Tsui, Y. H. Xie and
M. Shayegan, Europhys. Lett., \textbf{46}, 775 (1999)\ 

\bibitem{Jain} J. K. Jain, Phys. Rev. Lett., \textbf{63}, 199 (1989).

\bibitem{Tsui} W. Pan, H. L. Stormer, D.C. Tsui, L. N. Pfeiffer, K.W.
Baldwin, and K.W. West, Int. J. Mod. Phys. B \textbf{16}, 2940 (2002); Phys.
Rev. Lett. \textbf{90}, 016801 (2003).

\bibitem{Jain 04} C.-C Chang and J. K. Jain, Phys. Rev. Lett. \textbf{92},
196806 (2004).

\bibitem{Frad} Ana Lopez, and Eduardo Fradkin, Phys. Rev. B \textbf{69},
155322 (2004).

\bibitem{Go} M. O. Goerbig, P. Lederer, and C. Morais Smith, Phys.\ Rev.\ B 
\textbf{69}, 155324 (2004); Europhys. Lett. \textbf{68}, 72 (2004).

\bibitem{Wei} H. P. Wei, D. C. Tsui, and A. M. M. Pruisken, B 33, 1488
(1986).

\bibitem{M02} S. S. Murzin, M. Weiss, A. G. M. Jansen and K. Eberl, Phys.
Rev. B \textbf{66}, 233314 (2002).

\bibitem{Cl} R. G. Clark, J. R. Mallett, A. Usher, A. M. Suckling, R. J.
Nicholas, S, R. Haynes, and Y. Journaux, Surf. Sci. \textbf{196}, 219 (1988).

\bibitem{Sh} D. Shahar, D. C. Tsui, M. Shayegan, R. N. Bhatt, and J. E.
Cunningham, Phys.\ Rev.\ Lett.\ \textbf{74}, 4511 (1995).

\bibitem{Wong} L. W. Wong, H. W. Jiang, N. Trivedi, and E. Palm, Phys. Rev.
B \textbf{51}, 18033 (1995).

\bibitem{Dor2} S. I. Dorozhkin, M. O. Dorokhova, R. J. Haug, and K. Ploog,
Phys. Rev. B \textbf{55}, 4089 (1997).

\bibitem{Ch} T. Chakraborty, Surf. Sci. \textbf{229}, 16 (1990).

\bibitem{Kl} R. J. Haug, K. v. Klitzing, R. J. Nicholas, J. C. Maan, and G.
Weimann, Phys. Rev. B \textbf{33}, 4528 (1987)

\bibitem{Tsui87} B. E. Kane, D. C. Tsui, and G. Weimann, Phys. Rev. Lett. 
\textbf{59}, 1353 (1987).

\bibitem{Dor90} S. I. Dorozhkin, S. Koch, K. von Klitzing, and G. Dorda,
Pis'ma v ZhETF \textbf{52}, 1233 (1990) [JETP Lett. \textbf{52}, 652 (1990)].

\bibitem{Goldman} J.K. Wang and V.J. Goldman, Phys. Rev. Lett. \textbf{67},
749 (1991); Phys. Rev. B \textbf{45}, 13479 (1992).

\bibitem{Dor} S. I. Dorozhkin, G. V. Kravchenko, R. I. Haug, K. von
Klitzing, and K. Ploog, Pis'ma Zh.\ Eksp.\ Teor.\ Fiz. \textbf{58}, 893
(1993) [JETP Lett. \textbf{58}, 834 (1994)].

\bibitem{Eis} J. P. Eisenstein, L. N. Pfeiffer, and K. W. West, Phys. Rev. B 
\textbf{50}, 1760 (1994).

\bibitem{Rok} L. P. Rokhinson, B. Su, and V. J. Goldman, Phys. Rev. B 
\textbf{52}, 11 588 (1995).
\end{thebibliography}
\end{document}